# Estimation of hysteretic losses for $MgB_2$ tapes under the operating conditions of a generator

Carlos Roberto Vargas-Llanos[1] , Victor Zermeño[2] , Frederic Trillaud[3], Francesco Grilli[2]

[1] Posgrado en Ingeniería, National Autonomous University of México (UNAM), México

[2] Institute for Technical Physics, Karlsruhe Institute of Technology (KIT), Germany

[3] Institute of Engineering, National Autonomous University of México (UNAM), México

Email: varg1830@outlook.com

**Abstract**

Hysteretic losses in $MgB_2$ wound superconducting coils of a 500 kW synchronous hybrid generator were estimated as part of the European project SUPRAPOWER led by the Spanish company *Tecnalia Research and Innovation*. Particular interest was given to the losses found in tapes in the superconducting rotor caused by the magnetic flux ripples originating from the conventional stator during nominal operation. To compute the losses, a 2D Finite Element Method was applied to solve the *H*-formulation of Maxwell's equations considering the nonlinear properties of both the superconducting material and its surrounding Ni matrix. To be able to model all the different turns composing the winding of the superconducting rotor coils, three geometrical models of single tape cross section of decreasing complexity were studied: 1) the first model reproduced closely the actual cross section obtained from micrographs, 2) the second model was obtained from the computed elastoplastic deformation of a round Ni wire, 3) the last model was based on a simplified elliptic cross section. The last geometry allowed validating the modeling technique by comparing numerical losses with results from well-established analytical expressions. Additionally, the following cases of filament transpositions were studied: no, partial and full transposition. Finally, choosing the right level of geometrical details to predict the expected behavior of individual superconducting tapes in the rotor**,** the following operational regimes were studied: Bias-DC current, ramping current under ramping background field, and magnetic flux ripples under DC background current and field.

**Contents**

# 1.- Introduction

The incorporation of superconducting materials in electric machines presents various benefits such as size and weight reduction, greater efficiency at full and partial load, and higher power density [1], [2]. With such promises, superconducting rotatory machines emerge as an appealing technology to be applied to offshore multi-megawatt wind generators. Indeed, their greater power density can be used to maximize the "power per tower" by overcoming size and weight barriers of conventional turbines allowing thereby to reduce the overall cost associated with the construction, installation and maintenance of offshore wind towers [3], [4].

As wind power is expected to play a major role as one of the main renewable sources of energy in this decade, it is not surprising that various R&D projects on the design and operation of superconducting generators as well as the development and characterization of wires for wind power applications have been conducted worldwide considering the advantages that can be gained by such technology over conventional ones [5], [6], [7], [8], [9]. In Europe, the European Wind Energy Association has already predicted a need ranging from 165.6 GW to 217 GW of running wind power by 2020 to make up for decreasing one's dependence on conventional power generation and the constant increase of energy demand [10]. Greater effort is then necessary to meet the expectation and more research is and will be still necessary to overcome the technological challenges arising from the introduction of superconducting materials in wind power generation.

This present publication is part of a large R&D effort to characterize and develop $MgB_2$ wires for future mega-watt wind generators for offshore applications [11], [12], [13], [14]. Actually, the tape under analysis will be used in the SUPRAPOWER project [15], which aims to design and build a 10 MW hybrid synchronous generator for wind power applications. Each rotor pole of this machine will be composed of racetrack-shaped coils wound with copper laminated $MgB_2$ superconducting tapes whereas the stator will remain conventional. One prototype of 500 kW (4 rotor poles) is being built to validate some aspect of the design of the final mega-watt machine. One of the key aspects of the design is the ability to cool down and maintain the rotor to the desired operating temperature under different regimes of operation. To this end, a modular cryogen-free cooling system was developed and from an initial rough overall loss estimation, the average operating temperature is expected to be 20 K. In order to guarantee the expected operating temperature, a credible estimation of the steady state losses should be assessed. In this work, a particular focus is given on the AC losses induced in the superconducting tape triggered by the magnetic coupling between rotor and stator. Indeed, the magnetic ripples experienced by the rotor coils during the nominal operation of the generator induce dissipative currents in the superconducting windings that should be taken away by the cryogenic cooling system to ensure a reliable operation of the superconducting machine [16].

A large body of work has been dedicated to the estimation of losses in tapes and stacks. The first analyses relied on analytical expressions [17], [18], [19], [20] but soon proved to be too limited to specific problems and could be extended with difficuty to more practical general cases such as DC transport current under varying perpendicular magnetic field in stacks of multi-filamentary tapes. This complex case is particularly relevant when one deals with windings of thousands of turns typically encountered in superconducting electrical machines. To reach the level of geometrical details to study the behavior of multi-filamentary tapes in such stacks, new techniques based on the Finite Elements Method (FEM) were tried out. The method has quickly proven to be a powerful tool to analyze large scale applications through the use of homogenization and multi-scale based approaches [21], [22], [23], [24], [25].

To study the power losses in the superconducting rotor of a 500 kW synchronous wind generator, several considerations need to be made. So, the magnetic behavior of the Ni matrix was implemented through a nonlinear relative permeability, however its hysteretic losses were not considered. Furthermore, the coil ends were not modeled, in this way assuming that the machine is infinitely long along its rotatory axis. All the coils have the same number of turns and every turn is made of tapes of identical characteristics. The temperature of the superconducting winding is assumed homogeneous and equal to 20 K. Under these assumptions, a preliminary set of studies was carried out on a single tape to determine the basic geometrical details relevant to accurately compute the hysteretic losses in a stack using the multi-scale approach. Once the level of required details had been established, the superconducting rotor losses were computed for different operating conditions of the machine.

For single tape analysis, three models of tape cross section of decreasing complexity were studied: 1) the first model reproduced closely the actual tape cross section obtained from micrographs, 2) the second model was obtained from the computed elastoplastic deformation of a round Ni wire without considering the presence of $MgB_2$ filaments, 3) the last model was based on a simplified elliptic cross section which was used to validate the modeling technique through a comparison with well-established analytical expressions. To complete this study, three cases of filament transpositions are presented: no, partial and full transposition. This last study allows evaluating the impact of the transposition on the magnitude of the losses and determining the best practical transposition choice to manufacture tapes for wind power applications. Once the appropriate tape models were chosen to implement the multi-scale approach, the second analysis aimed at understanding the electric behavior of the superconducting tapes inside the rotor winding according to different regimes of the machine operation: 1) Bias-DC current, 2) ramping transport current with ramping DC background magnetic field, and 3) magnetic flux ripples on top of the DC background magnetic field with nominal DC transport current. After introduction the background of the study, results of the different studies leading to the multi-scale approach are discussed. Finally, an upper bound for the expected hysteretic losses in the superconducting rotor coils of the 500 kW prototype of the SUPRAPOWER project is given.

## 2.- Description of the hybrid synchronous prototype

The $MgB_2$ coils under study are part of a direct drive synchronous generator, developed in the SUPRAPOWER project led by the Spanish company *Tecnalia Research and Innovation* [26]. This project aims to design and construct a 10 MW hybrid machine for wind power application (superconducting rotor, conventional stator).

A first 500 kW prototype will be used to validate the 10 MW design. It is composed of 4 poles using the exact same coils per pole of the ones used in the final machine as depicted in Figure 1. Each superconducting coil is constituted of 9 Double Pancakes (DPs) separated by a G10 insulation layer. A DP is made of 2 pancakes wound with insulated multi-filamentary $MgB_2$ tapes, provided by *Columbus Superconductors*. The tapes are made of 19 $MgB_2$ filaments embedded in a Ni matrix and covered by a copper laminated layer on one side [27]. To help cooling and homogenizing the temperature of the DP, a copper plate is inserted between the pancakes. Table 1 summarizes the salient parameters of a pole coil and its cross section is shown in Figure 2.

**Table 1: Salient parameters of MgB2 tape and the superconducting coils**

| Parameter | Value |
|---|---|
| Number of pancakes per coil | 18 |
| Number of turns per pancake | 75 |
| Number of turns per coil | 1350 |
| Total wire length, in the coil | 3200 m |
| Critical current of the tape at 20K and 1.8 T | 158.07 A |
| Wire dimensions (bare) | 3 x 0.7 mm |
| Wire dimensions (insulated) | 3.125 x 0.825 mm |
| Number of filaments in tape | 19 |

## 3.- Methodology for estimating the hysteretic losses in a superconducting rotor coil

Finite Element Analysis has proven to be a valuable tool to estimate losses in single tapes and stacks of tapes. However, it is still a difficult task especially if one must consider some degree of geometrical details of tape cross sections. Thus, implementing multi-filamentary tapes is a challenge that requires high computing resources and time.

In the present case, a preliminary set of studies are conducted on a single tape for which the impact of a detailed geometrical description of the tape cross section and the filament transpositions on AC losses is investigated. These preliminary studies allow determining the best level of geometrical details of tape cross section to compute with the best possible accuracy the AC losses over the whole rotor coils. These losses, hysteretic in nature, originate from magnetic flux ripples coming from the stator slots. Thus, in a second phase before simulating the whole coil, the impact of these ripples on a single tape is studied. First, the simulation of a DC current ramp and a magnetic field applied perpendicularly to the tape surface are simulated representing the loading regime of the coil. Once the nominal transport current in the coil is reached and the transient behavior passed, magnetic ripples of different amplitudes are superimposed to the perpendicular magnetic field, and the average hysteretic losses of a single tape are then computed.

Based on the single tape analysis, it was found that the simulation of the superconducting coil cross section modeling each individual filament making a tape is not practical, so a multi-scale approach was used as proposed in [24], [23], [28]. Hence, the magnetic field at the coil boundary is split into a DC component and magnetic ripples using Fast Fourier Transform (FFT). For the computation of the DC component, the model uses the elliptic simplification of the tape cross section. The magnetic field at each tape boundary computed with the previous simplified model is then used to estimate the losses in a single tape resulting from the magnetic ripples for which a greater level of details, including filaments and their transposition, is considered. Finally, an upper bound of losses can be inferred using the multi-scale approach. This approach allows the estimation of the 2D distribution of losses in coil cross section showing areas of higher and lower heat dissipations that can be fed to the thermal analysis of the cryogenic system.

## 4.- Individual tape cross sections and transpositions

To compute AC transport current losses, three cross sections are considered. The first one is referred to as original cross section (Figure 4, top model). It is digitalized from a tape micrograph. One important feature of this geometry is the relative proportion of materials (25.32% Ni; 74.68% $MgB_2$) that does not match the proportions provided by the manufacture (21.5% Ni; 78.5% $MgB_2$) [27]. Therefore, a second standard geometry was developed, referred to as standard cross section. Thus, the actual tape cross section was computed from the elastoplastic deformation of a nickel cylinder without considering the presence of the $MgB_2$ filaments as shown in Figure 5 [15]. Once the deformation was computed keeping the material proportions as provided by the manufacturer specifications, the copper layer was added on one side of the tape to obtain the desired tape model as given in Figure 4 (middle model). Finally, an additional elliptic cross section, referred to as elliptic cross section, was considered. It is shown in Figure 4 (bottom model). This final simplified cross section allowed comparing the numerical results to the results obtained from well-established analytical formulas to validate the modeling technique. To match the measured critical current with the one expected from the elliptic model, the $J_c(B)$ expression was multiplied by a scaling factor ($F_r = 0.464$). The resulting mismatch is lower than 1% over the whole range of measurements. To compute the critical current the methodology explained in [29] was implemented.

Fields changing in time tends to couple filaments together within individual wires and to couple wires together within cables. To reduce this effect, which increases AC losses, both wires and cables should be transposed [30]. Bearing in mind the manufacturing technique to produce the tape, three transposition cases were considered [31]. A full transposition case, all the filaments exchange position (which is not possible in real samples); so all the filaments are carrying the same amount of current. A partial transposition case, the filaments are grouped in three layers (external layer: 12 filaments, middle layer: 6 filaments and central layer: 1 filament) where only the filaments in the same layer exchange position. In this case, each filament of each layer transports the same amount of current but the current differs from one layer to the other. For the no transposition case, none of the filaments exchanges position, so each filament could carry any amounts of current.

As a result, 7 cases ought to be studied: original cross section with full, partial and no transposition; standard cross section with full, partial and no transposition; and finally elliptic cross section for which no transposition applies. Before presenting the numerical results, a revision of the mathematical equations and the basic modeling technique is briefly presented.

## 5.- Mathematical equations and modeling technique

### 5.1. Critical current density and E-J power law

To describe the macroscopic behavior of the superconducting material a non-linear $\vec{E} - \vec{J}$ relation is used. This relation is expressed in Equation 1 as follows

$$\vec{E} = Ec \left|\frac{\vec{J}}{J_c(\vec{B})}\right|^{n-1} \frac{\vec{J}}{J_c(\vec{B})} \qquad \text{Equation 1}$$

Where $Ec$ corresponds to the critical electric field ($Ec = 10^{-4} V/m$), $J_c$ is the critical current density, $\vec{B}$ is the magnetic flux density and $n$ is the $n$-value indicating the sharpness of the *E-J* characteristic [32]. Equation 2 gives the expression of the magnetic dependence of critical current density as follows [1],

$$J_c(B) = J_{C0}\left(1 - \frac{B}{B_1}\right)\left(1 + \frac{B}{B_0}\right)^{-\alpha} \quad \text{Equation 2}$$

The best fit for the experimental data of [27] was obtained with the parameters summarized in Table 2 [29] Figure 3. The relative error lower than 1% over the whole range of measurements.

**Table 2: Values of parameters for best fit of critical current density**

| Parameter | Value |
|---|---|
| $J_{C0}$ | $3.4874 \times 10^9 A/m^2$ |
| $B_0$ | $0.034469\ T$ |
| $B_1$ | $3.1213\ T$ |
| $\alpha$ | $0.25$ |
| $I_C$ | $649.24$ A |

### 5.2. *H*-formulation

Computation of losses was conducted using Comsol Multiphysics® implementing a 2D *H*-formulation. To take into account the magnetic property of the Ni matrix embedding the MgB$_2$ filaments, Maxwell-Ampére and Maxwell-Faraday equations together with the constitutive relations were expanded with a nonlinear relative permeability depending on the applied magnetic field *H* as described in [22]. The resulting system for the 2D *H*-formulation is presented in equations (Equation 3), (Equation 4) and (Equation 5) [22]. Where $H_x$ and $H_y$ are the *x* and *y* components of the magnetic field, $E_z$ is the *z* component of the electric field, $J_z$ is the *z* component of the current density, $\mu_0$ is the vacuum permeability and $\mu_r(H)$ corresponds to the relative permeability depending on magnetic field (Figure 6). Finally, the resistivity of the superconducting material is given by equation (Equation 6). The resistivity of copper ($\rho_{Cu}|_{20\ K} = 0.028 \times 10^{-9} \Omega \cdot m$) and nickel ($\rho_{Ni}|_{20\ K} = 0.014 \times 10^{-8} \Omega \cdot m$) was estimated at an operating temperature of 20 K [33].

$$\frac{\partial E_z}{\partial x} = \mu_0 \left[ \frac{\partial \mu_r(H)}{\partial t} H_y + \mu_r(H) \frac{\partial H_y}{\partial t} \right]$$    Equation 3

$$\frac{\partial E_z}{\partial y} = -\mu_0 \left[ \frac{\partial \mu_r(H)}{\partial t} H_x + \mu_r(H) \frac{\partial H_x}{\partial t} \right]$$    Equation 4

$$J_z = \frac{\partial H_x}{\partial y} - \frac{\partial H_y}{\partial x}$$    Equation 5

$$\rho_S = \left| \frac{\vec{J}}{J_c(\vec{B})} \right|^{n(B)-1} \frac{Ec}{J_c(\vec{B})}$$    Equation 6

To compute of the instantaneous losses, the dot product between the electric field $\vec{E}$ and the current density $\vec{J}$ was computed over the whole cross section of the tape. The average losses ($Q$) were obtained by integrating the instantaneous losses over half a period ($T$) multiplying the result by a factor $2/T$ as shown in (Equation 7).

$$Q = \frac{2}{T} \int_{T/2}^{T} \left[ \int_S \vec{E} \cdot \vec{J} ds \right] dt$$    Equation 7

## 6.- Individual tape analysis

### 6.1.- Verification of the modeling technique

To validate the modeling technique, the losses due to AC transport current obtained with FEM were compared with an analytical equation developed by W.T. Norris [18]. The amplitude of the AC transport current was spanned from 10% $I_c$ to 100% $I_c$ as shown in Figure 7.

Having cross-checked the modeling technique with a well-established analytical expression in the simplest assumptions, the full model was studied. Thus, the total AC losses for the three cross sections (original, standard and elliptic) and the transposition cases were computed with AC transport current having amplitudes ranging from 10% $I_c$ to 100% $I_c$. This study is part of understanding the impact of the manufacturing process of tapes on AC losses. Indeed, the fabrication of twisted filaments is a complicated task since breaks and filament joints may lower the performance of the tape for practical operations. The results are gathered in Figure 8. The losses estimated using standard cross section are shown using a diamonds symbols, elliptic cross section using rectangular symbols and original cross section using circles. Each transposition case was identified through different line types (continuous: no transposition, dashed: partial, dots: full). It can be noted in Figure 8 an increase of average losses due to the saturation of some of the tape filaments for currents ranging from 70% to 100% of $I_c$ in the no transposition case (continuous line). The outermost filaments shielding the penetration of the magnetic field carry most of the current. A similar saturation effect could be seen for the partial transposition (dashed line). However it is weaker than the previous case because of the uniformization of current distribution. Similarly it can be noted that there is a discrepancy between average losses of standard (diamonds symbols) and original (circles symbols) geometrical models which originates from the difference in material proportions and filament shapes.

The transposition of filaments in the superconducting tape plays a significant role in reducing the losses at transport current reaching the critical current. However, it tends to be miscellaneous at low transport currents. Amongst the different cases of filament transposition, the only model representing the actual tape to be used in the SUPRAPOWER project is the partial transposition case [31]. For this reason it will be used for subsequent analysis. In practice, the full transposition case is difficult to be implemented in the manufacturing process of the tapes even though it may improve the average losses at higher values of transport current. It should be noted that the improvement in losses compared to the partial transposition case is less than a factor of 2 at $I_c$ and this difference decreases quickly at lower fraction of $I_c$ (< 80%) at which HTS systems are usually operated. It is then the optimum manufacturing process to meet the operating specification of the generator. The no transposition case is simpler but only gives an upper limit to the average losses. It is not the preferred choice of tape manufacturing process since it does not make use of most of the superconducting material dictating the cost of the product.

### 6.2.- Estimation of average tape losses due to DC transport current and DC background magnetic field with and without ripples

This analysis is conducted in two steps. A first step introduces a 100 s magnetic ramp from 0 T to the operating value while a current is ramped from 0 A to 95 A (nominal current of the superconducting coils). The external magnetic field is applied perpendicularly to the tape surface. Once most of the transient regime has vanished (> 600 s as depicted in Figure 9), the last state of the previous step (at 1000 s in Figure 9) is transferred to the next step as an initial condition. In this second step, ripples are added to the perpendicular magnetic flux density at a transport current of 95 A. These ripples are sinusoidal functions of amplitudes: 1 mT, 2 mT, 3 mT, 4 mT, 10 mT and 100 mT and frequency of 24.3 Hz. This frequency is set to match the 6$^{th}$ harmonic of the magnetic field generated in the generator (see next section). Since the initial conditions for each ripple amplitude is identical, the first step is computed only once at the same external field and DC transport current.

The instantaneous losses during the ramping period (1st step) are shown in Figure 9, where one can note the increase of losses during the ramping to reach a maximum before decaying over the first hundreds of seconds before slowing. This transient behavior is related to the induction of persistent currents during the ramping period [34]. These induced currents decay slowly due to the low resistivity of the $MgB_2$ filaments modeled here as a power law. The average losses per unit length as a function of ripple amplitudes are shown in Figure 10. It can be noticed average loss increase with ripples amplitude. Indeed, when the ripple amplitudes are increased, the area under hysteresis loop gets bigger leading to an increase in the losses.

From these last results and assuming that all the tapes of the superconducting rotor coils behave accordingly to a single tape under magnetic ripples, an upper limit of the hysteretic losses in the rotor of the 500 kW machine is subsequently estimated in section 7.2.

## 7.- Analysis of the superconducting rotor coils

### 7.1.- Magnetic flux density in the rotor coil

Knowing the value of magnetic field density at the boundary of the cross section of one straight segment of a rotor coil, the distribution of the magnetic flux density inside the coil is computed on the basis of the elliptic model of individual tape cross section (Figure 12). This boundary covering the path traced by point 1 to 4 is given in Figure 11. It is divided in two parts: DC and ripples. Knowing the distribution of the magnetic field inside the coil, it is possible to extract the value of the local magnetic flux density at the boundary of each tape. This value is then used as boundary condition for a model considering a single tape for which the losses can be computed considering a greater level of geometrical details of the tape such as the standard cross section presented in single tape analysis. From the distribution of the magnetic flux density given in Figure 12, one can observe that the strongest magnetic field is located at the upper left side of the coil, and the lowest intensity is at the lower right side.

Using FFT to analyze the magnetic ripples resulting from the coupling between rotor and stator, the amplitude spectrum was computed and shown in Figure 13 and Figure 14. Considering an operating frequency of 4.05 Hz, even harmonics are present with the $6^{th}$ one showing the largest amplitude (24.3 Hz) almost all along the boundary of the coil (path 1 to 4). This harmonic is used to compute the hysteretic losses.

### 7.2.- Estimation of coil loss

For the tape in the superconducting coil exposed to a magnetic flux density of 1 T with a ripple amplitude of 1 mT at a DC transport current of 95 A, the average loss is 0.12 mW/m which correspond to an upper limit for the remaining tapes in the coil. For a total tape length in the coil of 3200 m [15], the total loss per coil is 0.4 W. For the prototype, considering 4 coils, the burden on the cryogenic system is 1.6 W well below the cooling power of commercial cryocoolers. Even though a better estimation could be achieved by taking into account the losses of each individual tape of the coil, the upper bound is the most relevant parameter to ensure safe operation of the machine.

For completeness and to get a range of expected loss values for such a generator, a multi-scale technique was implemented to estimate losses in the superconducting coil. The DC component of the magnetic field was obtained from a magnetostatic model of the full coil. To compute the magnetic field, the elliptic cross section model was used since it requires less amount of elements and gives a good representation of magnetic field outside the tape. Then, the computed DC magnetic field for which magnetic ripples were added was used as boundary condition to study the losses of each individual tape of the coil cross section. The individual tape was modeled following the standard geometry with partial transposition. In Figure 15 a representation of this strategy can be seen, where the boundary of the tape was depicted with a continuous blue line. The amplitude of the magnetic ripples was chosen based on the inspection of Figure 16 showing the quotient between the average ripple magnitude and the DC component at the boundary of the coil. Hence, for the tapes in the uppermost and lowermost pancakes of the coil, a mean value of magnetic ripples was used as boundary condition (upper path 1-2 and lower path 3-4). For the rest of the tapes, a linear interpolation was applied. This will overestimate losses, since the ripples are not considered affected by the currents induced in the metallic materials. In this study, average tape losses were calculated over 40 tapes of the 75 making each pancake (720 out of 1350 tapes**)**. The remaining tape losses were estimated by interpolating the average losses of the 40 tapes per pancake over the entire coil cross section. In this study, the total losses per coil were estimated equal to 0.01 W, one order of magnitude lower than the upper bound as previously computed. Figure 17 shows the average tape losses as function of the tape location. Tapes were numbered from left to right. Each pancake was highlighted with a different color and numbered from bottom to top. The tapes with greater losses are located in the uppermost pancake for which extra care would have to be taken to maintain the operating temperature at an average of 20 K. Figure 18 shows the distribution of losses in the coil cross section. A strong relation between average losses per tape and the DC background magnetic field can be inferred if one compares Figure 18 with Figure 12. Here the highest losses and the strongest magnetic field are found at the top of the coil. Conversely, lower losses and the weakest magnetic field can be found at the lower right corner of the coil. It is expected, as the nonlinear resistivity of the superconductor responsible for the hysteretic losses is strongly dependent on the local value of the magnetic flux density. Indeed, greater magnetic field leads to larger resistance. From the distribution of losses, it may be inferred that the uppermost pancakes would require better cooling.

## 8.- Conclusions

The hysteretic losses of multi-filamentary MgB$_2$ tapes under synchronous generator operating condition were estimated by means of FEM and *H*-formulation considering a certain level of geometrical details. The following operating conditions were analyzed: AC transport current, DC transport current under external applied magnetic field with and without ripples. The validation of the modeling technique was carried out estimating losses with the simplest elliptic cross section model. However, greater geometrical details including partial filament transposition were implemented to get a better estimation of the average tape losses. Therefore, the Cu layer and the Ni matrix as well as more realistic operating conditions could be taken into account. Based on the chosen geometrical tape model, DC transport current and uniform applied magnetic field losses were estimated on single tapes. Adding magnetic ripples to include the coupling between rotor and stator, it was found that the average tape losses increased rapidly with increasing ripple amplitude, as expected from the hysteretic nature of losses in superconducting materials.

Finally, the coil losses could be estimated using a multi-scale approach. An upper limit of 0.4 W and a lower limit of 0.01 W per coil were estimated to give a range of values to design safety the cryogenic cooling system thereby ensuring the reliable operation of the generator. A strong dependence was found between the background DC magnetic field and the average losses of the superconducting coil. The greatest losses were found at tapes under the strongest DC background magnetic field.

## 9.- Acknowledgement

The authors thank Tecnalia Research & Innovation for providing useful data. First author thanks Coordinación de Estudios de Posgrado (CEP-UNAM) and Consejo Nacional de Ciencia y Tecnología (CONACYT) for financial support. Second and fourth authors acknowledge the support of the Helmholtz Association (Young Investigator Group grant VH-NG-617).

## 10.- References

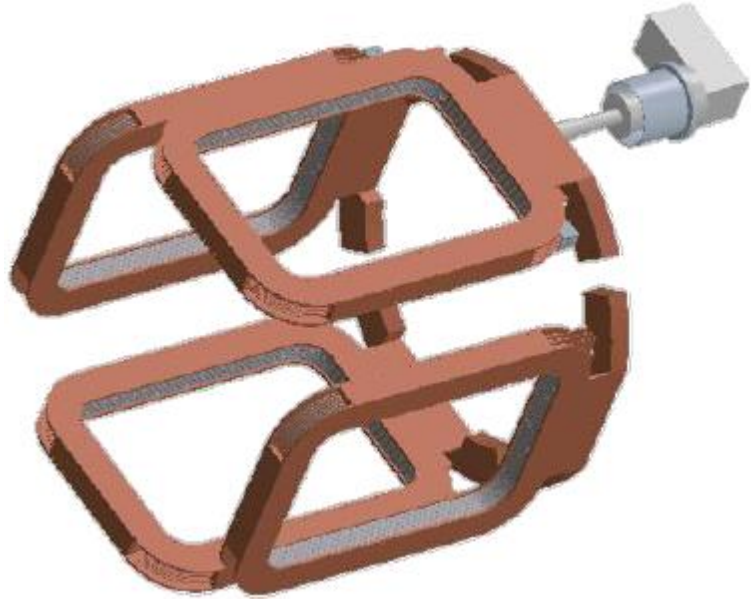

**Figure 1: 3D CAD model of the 4-pole superconducting rotor of the 500 kW SUPRAPOWER prototype [35]**

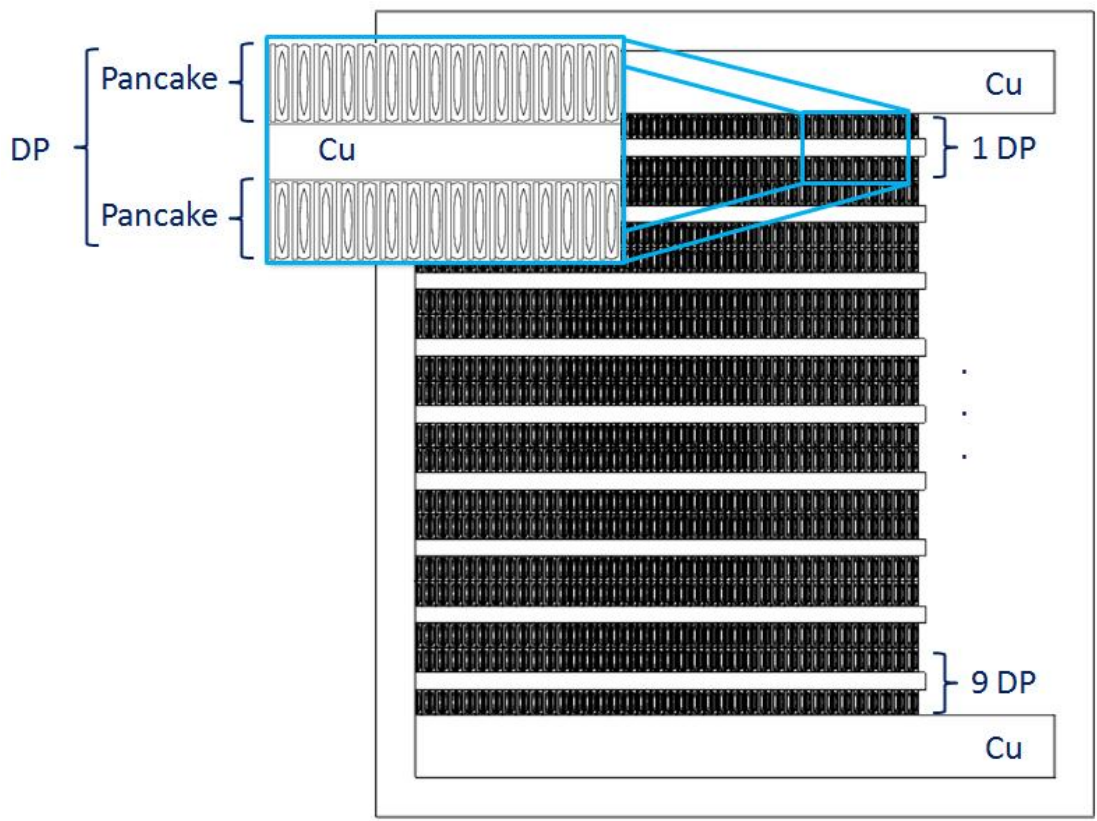

**Figure 2: Detailed cross section of the superconducting rotor coil**

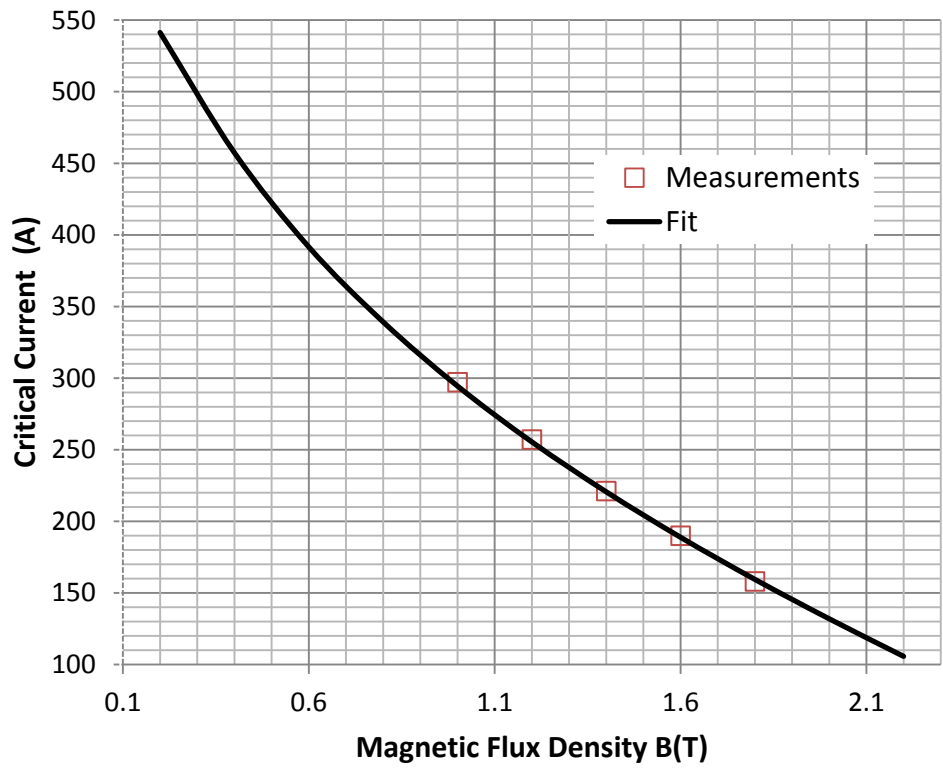

**Figure 3: Fit of the critical current**

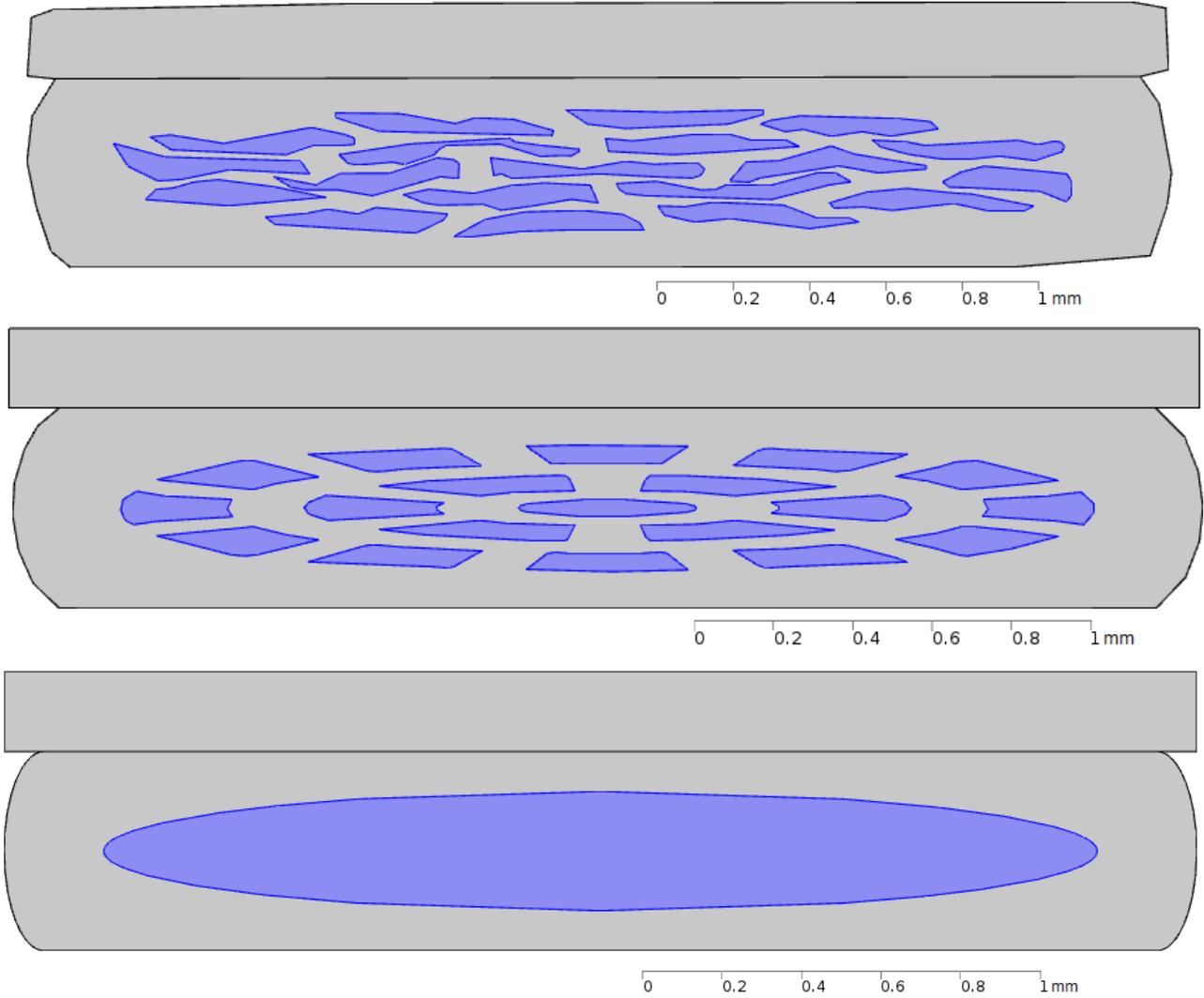

**Figure 4: From top to bottom: original, standard and elliptic tape cross section model**

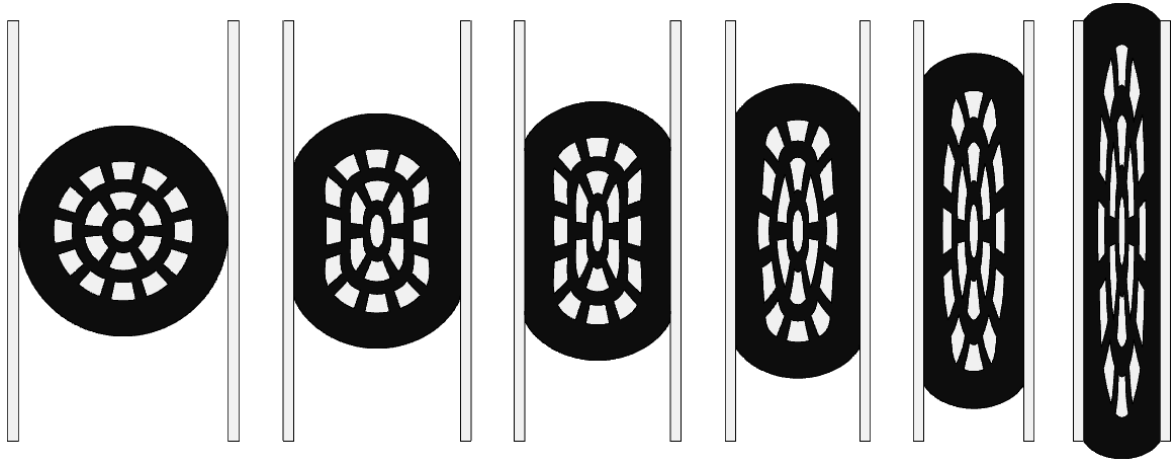

**Figure 5: Elastoplastic deformation of nickel cylinder, without considering the presence of the MgB$_2$ filaments [36]**

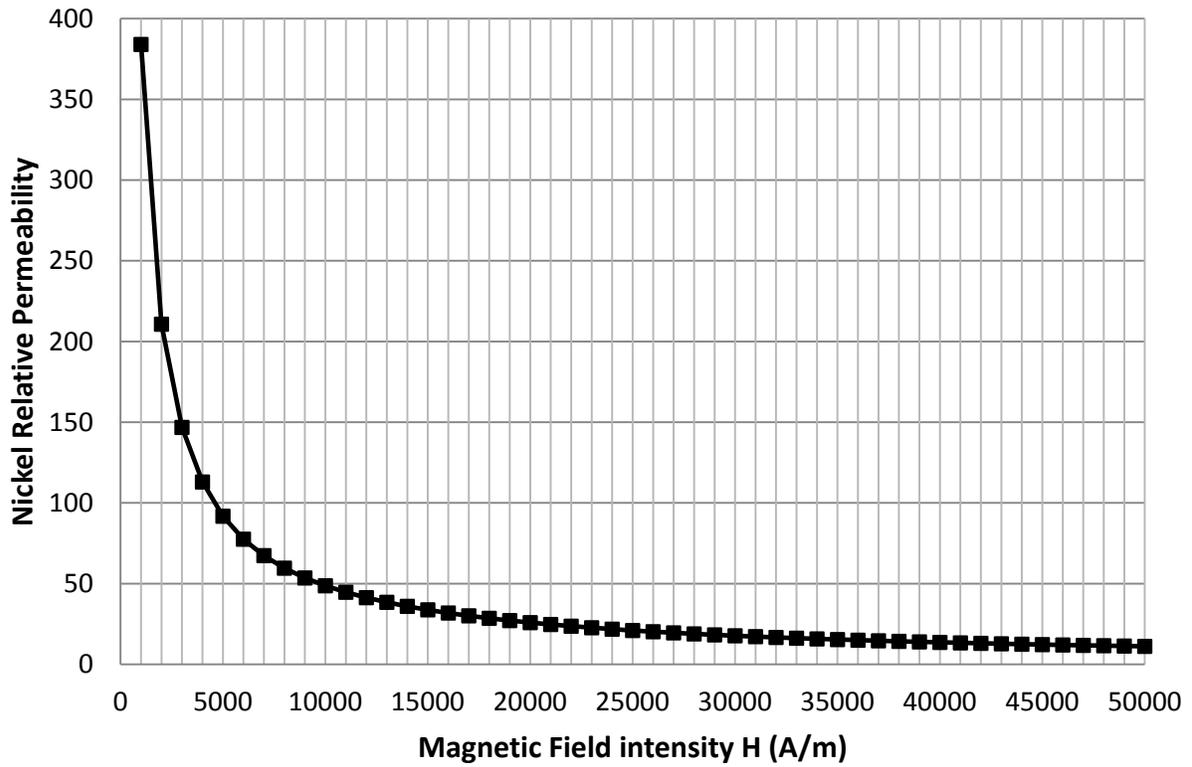

**Figure 6: Nickel relative permeability as a function of magnetic field intensity**

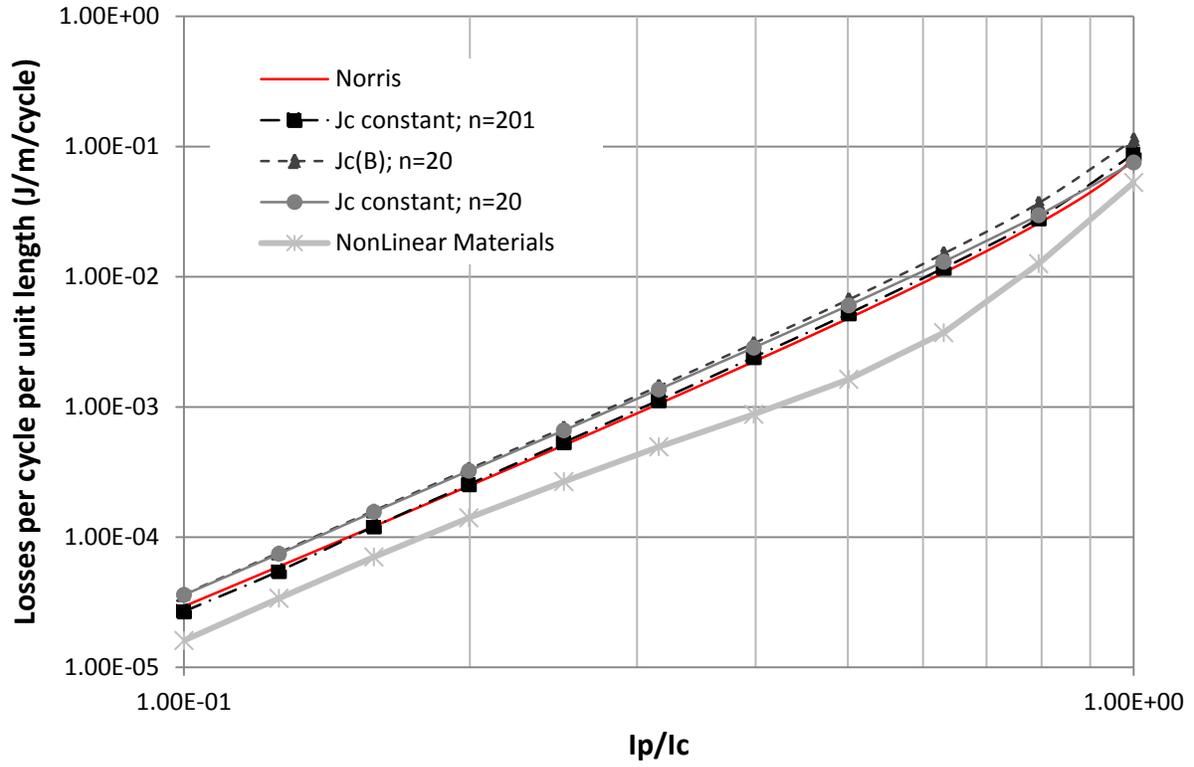

**Figure 7: Comparison of losses in MgB$_2$ tapes using the proposed model with W.T. Norris' analytical equation[18]**

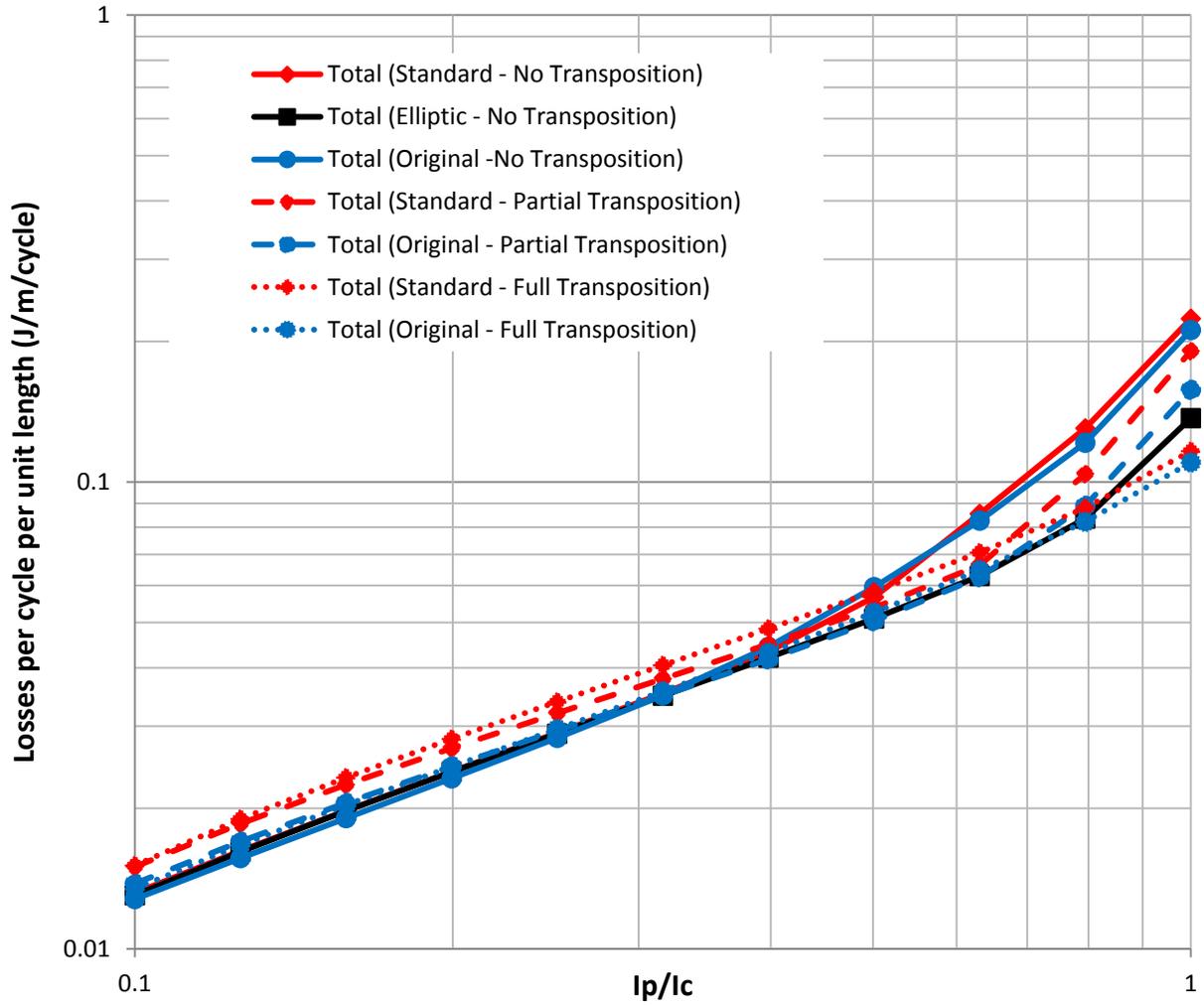

**Figure 8: Total losses due to AC transport current without applied external magnetic field, for the original, standard and elliptic cross section considering full, partial and no transposition**

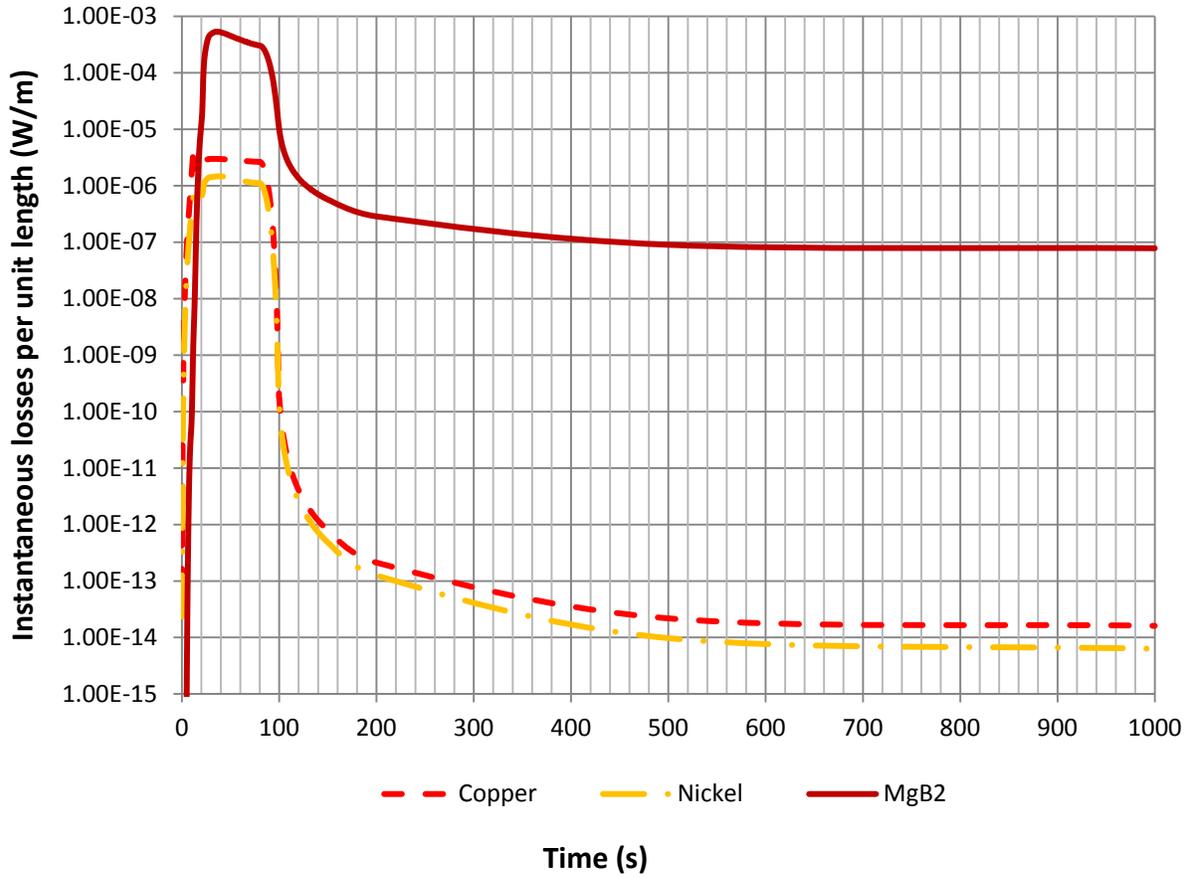

**Figure 9: Instantaneous losses for each material constituting the tape during current and DC background magnetic field ramping, for a operating current of 95 A and a background magnetic field of 1 T**

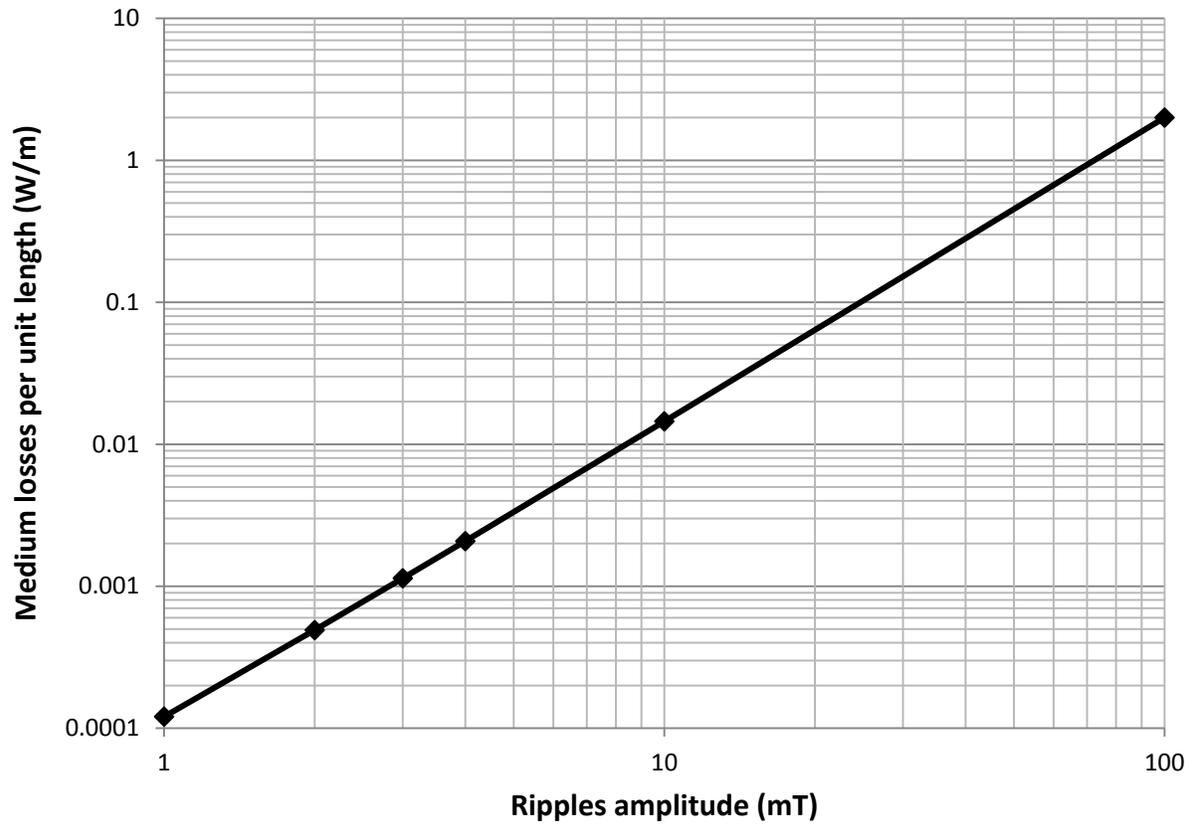

**Figure 10: Average tape losses under a transport current of 95 A and 1 T DC background magnetic field perpendicular to the tape, as a function of the amplitude of the magnetic ripples**

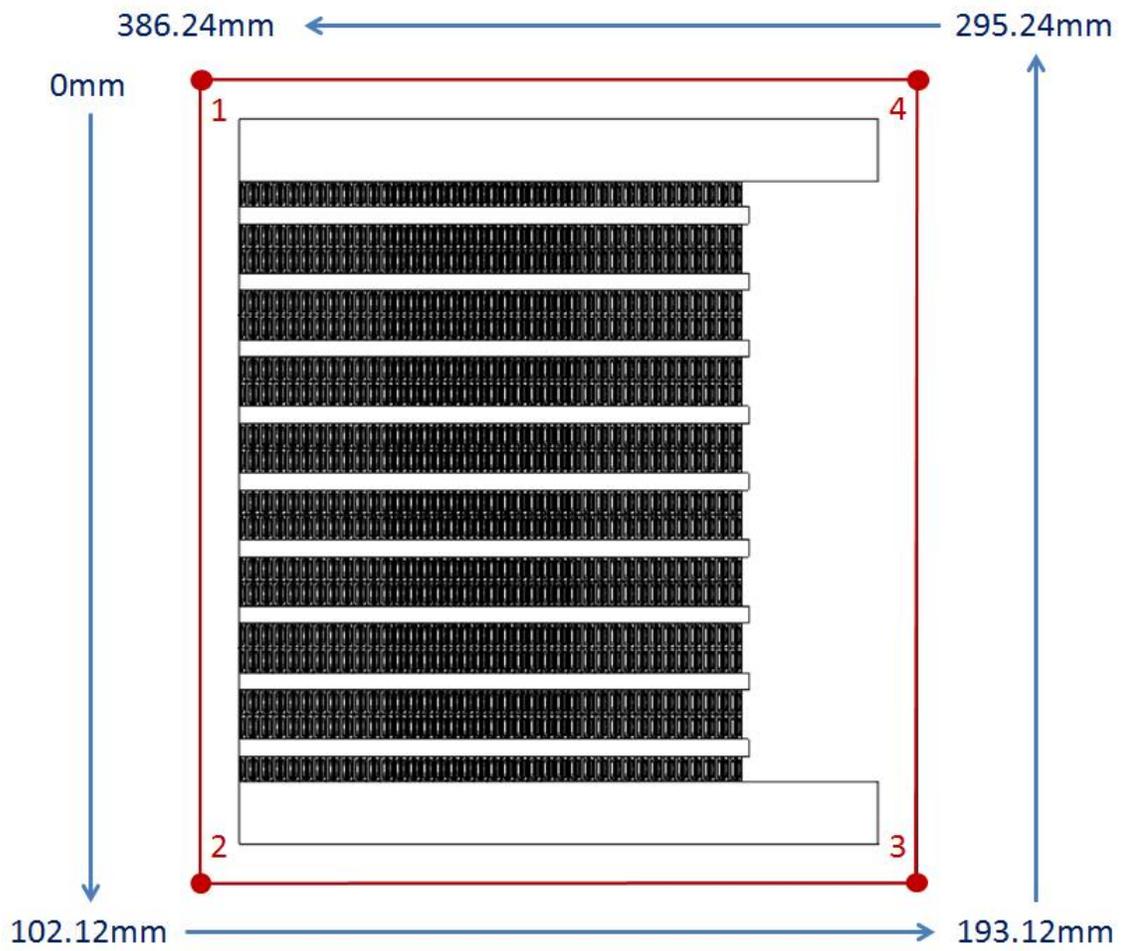

**Figure 11: Rectangular boundary of the superconducting coil covering the path passing by points 1 to 4**

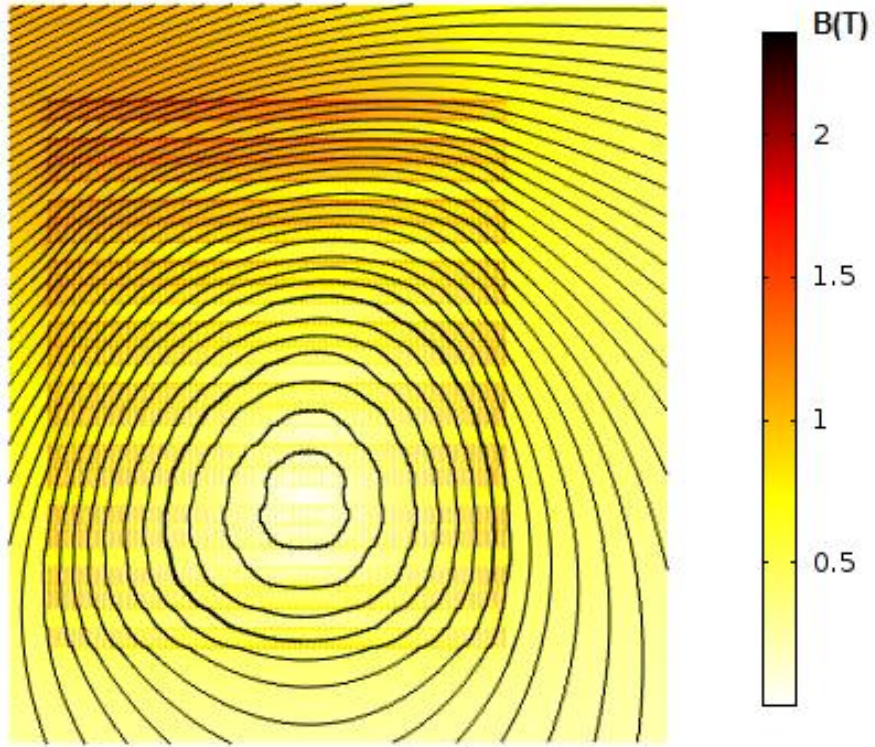

**Figure 12: DC Magnetic field in the cross section of the rotor coil, without taking into account the magnetic ripples**

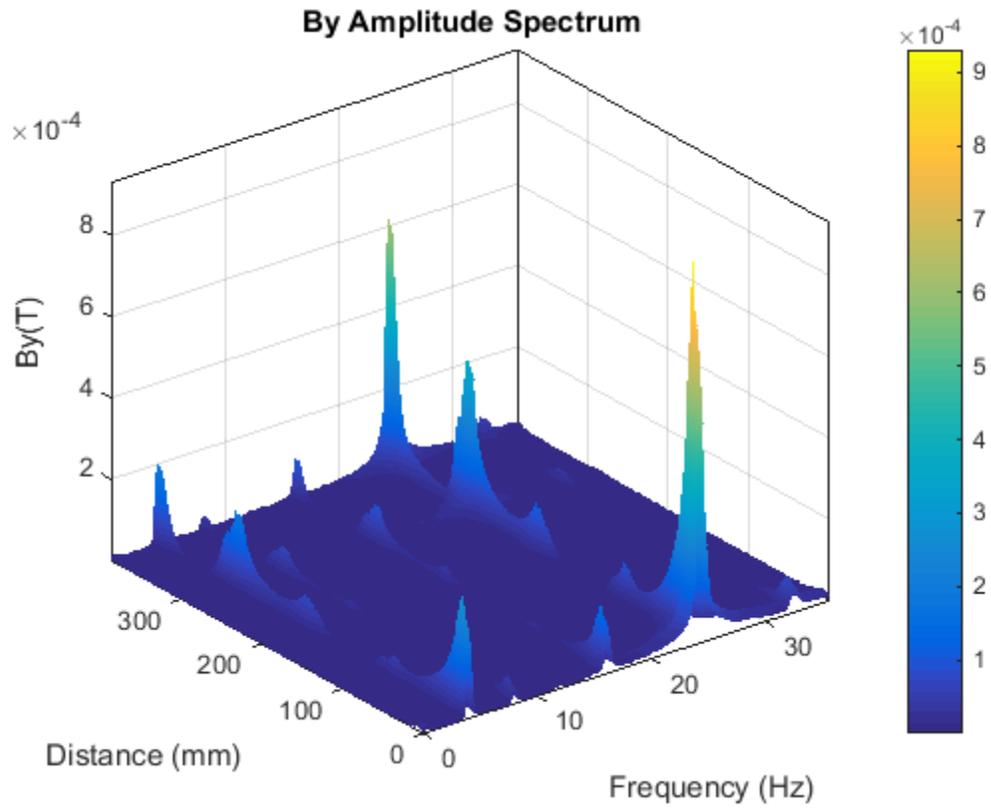

**Figure 13: Amplitude spectrum of the *y* component of the magnetic field at the coil boundary (see Figure 11)**

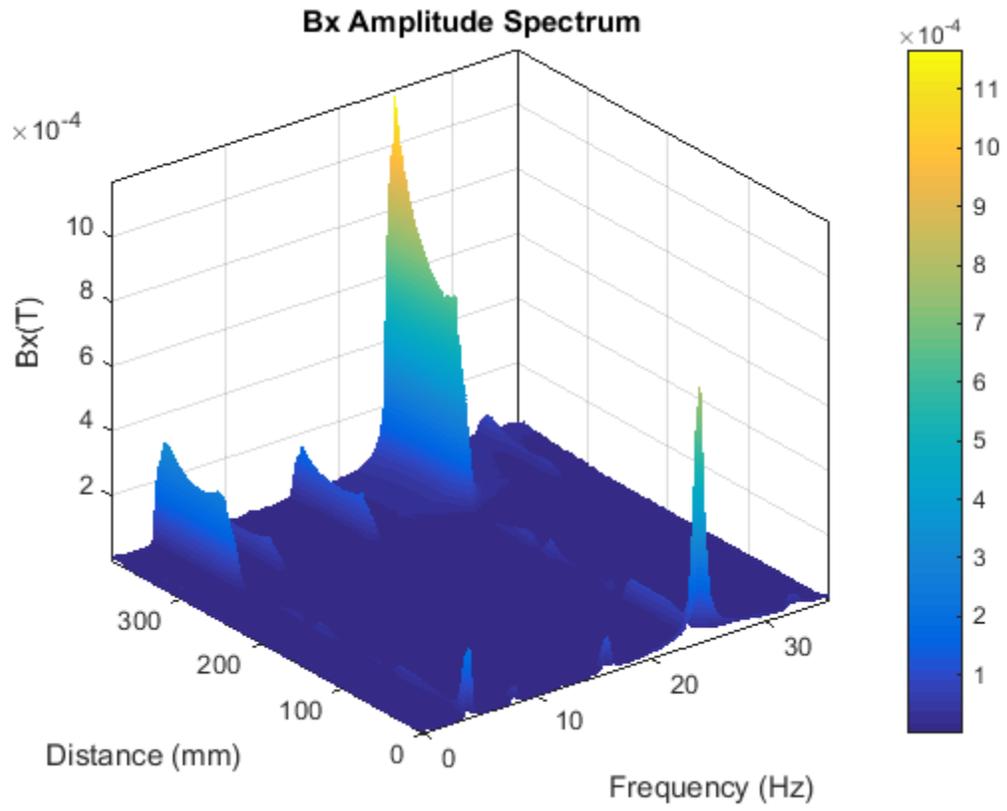

**Figure 14: Amplitude spectrum of the *x* component of the magnetic field at the coil boundary**

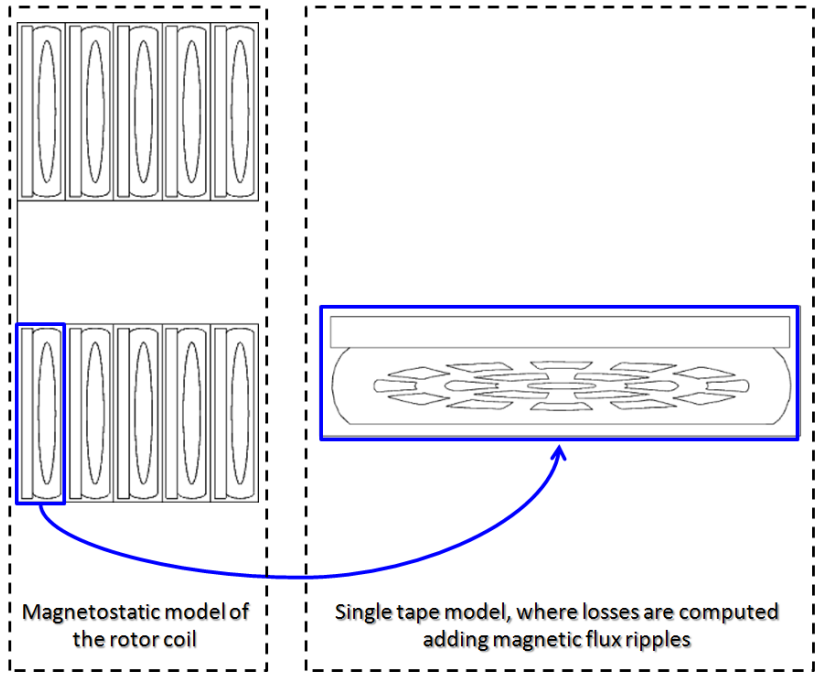

**Figure 15: Multi-scale strategy implemented for the estimation of losses in the superconducting coil**

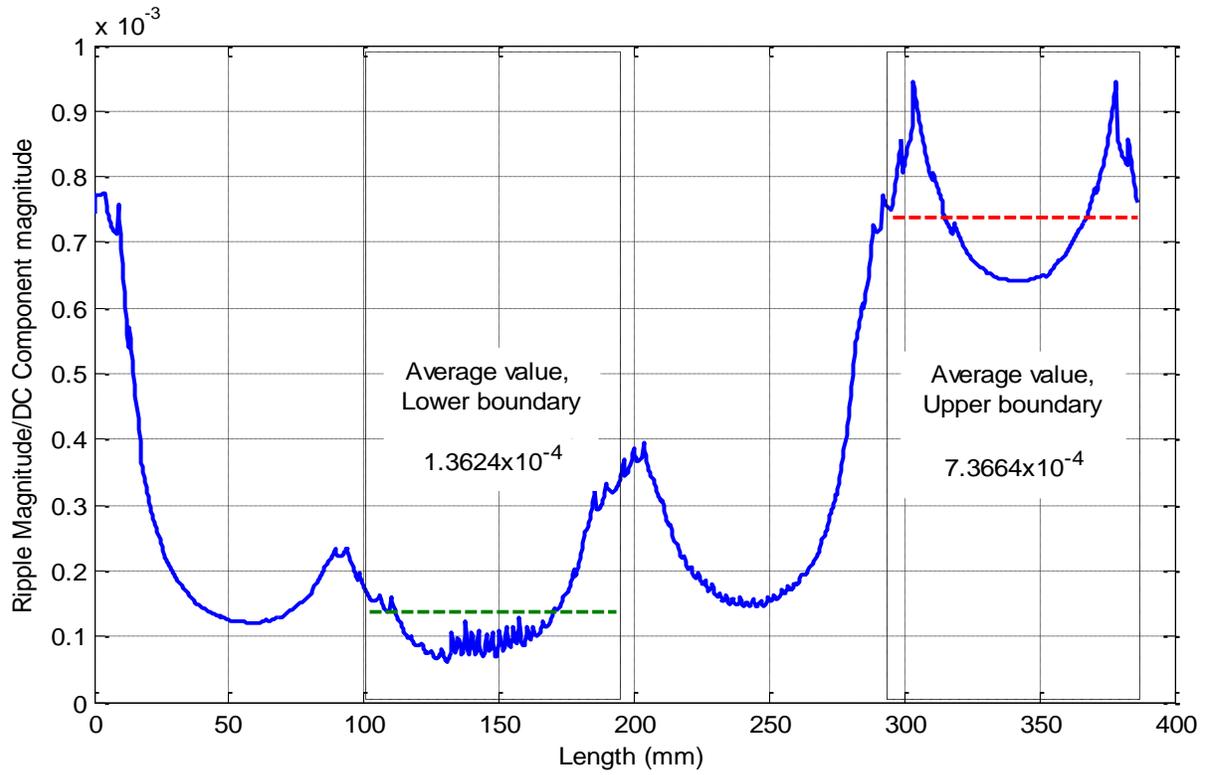

**Figure 16: Quotient between the magnetic flux ripples and the DC component of the magnetic field at the boundary as a function of path length from point 1 to 4 (see Figure 11)**

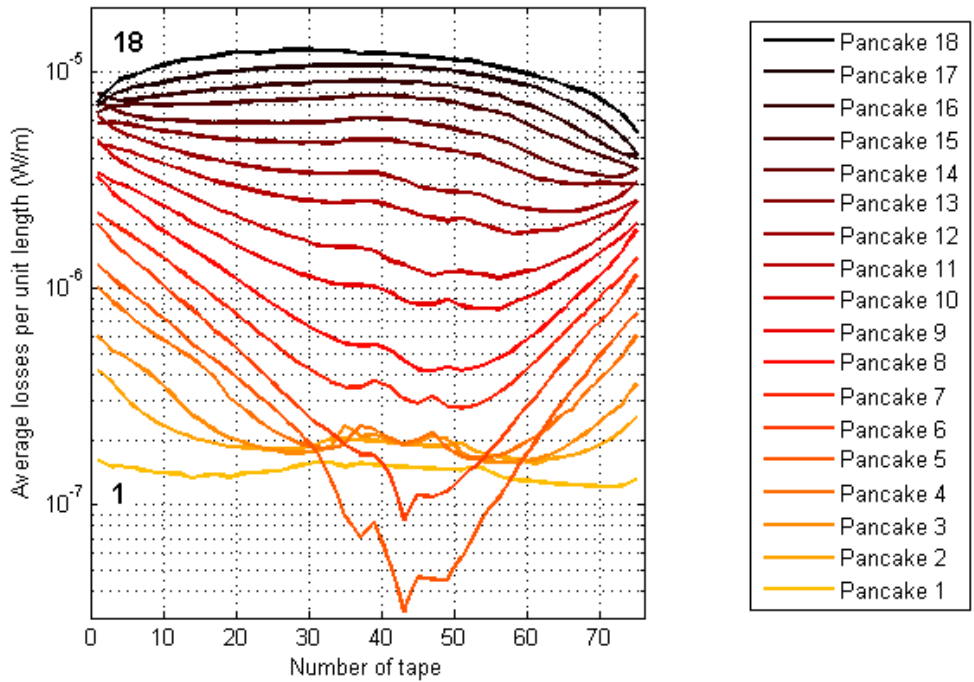

**Figure 17: Average losses per pancake as a function of the number of tapes**

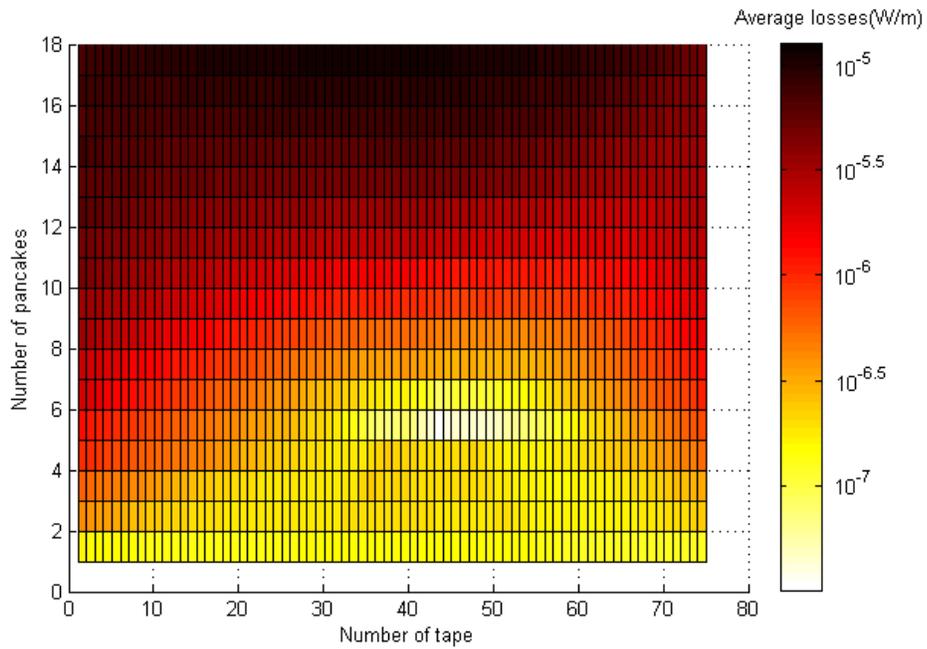

**Figure 18: Distribution of the average tape losses in the coil**